\documentclass[a4paper,12pt]{article}
\newcommand{\nin}{\noindent}
\newcommand{\be}{\begin{equation}}
\newcommand{\ee}{\end{equation}}
\newcommand{\bea}{\begin{eqnarray}}
\newcommand{\eea}{\end{eqnarray}}
\newcommand{\nn}{\nonumber\\}
\newcommand{\tx}{\bar{X}}
\newcommand{\olo}{\int d^2}

\newcommand{\la}{\Lambda^2}

\newcommand{\miso}{\frac{1}{2}}

\newcommand{\ap}{\alpha'}

\usepackage{graphicx}

\begin{document}

\begin{titlepage}
\begin{center}

{\bf{\Large Tachyon-Dilaton-induced Inflation as an $\alpha^{'}$-resummed String Background}}

\vspace{1cm}

\textbf{Jean Alexandre\footnote{jean.alexandre@kcl.ac.uk}, Anna Kostouki\footnote{anna.kostouki@kcl.ac.uk}} and
\textbf{Nick E. Mavromatos\footnote{nikolaos.mavromatos@kcl.ac.uk}\\}

\vspace{0.5cm}

Department of Physics, Kings's College London, \\
London WC2R 2LS, U.K.

\vspace{0.5cm}

{\bf Abstact}

\end{center}

\vspace{0.2cm}

Within the framework of a novel functional method on the world-sheet of the string, we
discuss simple but re-summed (in the Regge slope) inflationary scenarios in the context of closed Bosonic strings, living in four target-space dimensions, in the presence of non-trivial
tachyon, dilaton and graviton cosmological backgrounds. The inflationary solutions are argued to guarantee the vanishing of the corresponding Weyl anomaly coefficients in a given world-sheet renormalization scheme, thereby  ensuring conformal invariance of the corresponding $\sigma$-model to all orders in the Regge slope. The key property is the requirement of ``homogeneity'' of the corresponding Weyl anomaly coefficients. Inflation entails appropriate relations between the dilaton and tachyon field configurations, whose form can lead to either a de Sitter vacuum, incompatible though (due to the cosmic horizons) with the perturbative string scattering amplitudes, or to cosmic space-times involving brief inflationary periods, interpolating smoothly between power-law and/or Minkowski Universes. The latter situation is characterized by  well-defined scattering amplitudes, and is thus compatible with
a perturbative string framework. It is this scenario that we consider a self-consistent ground state
in our framework, which is based on local field redefinitions of background fields.

\vspace{2.5cm}
\begin{flushleft}
November 2008
\end{flushleft}

\end{titlepage}

\section{Introduction and Summary}

The inflationary paradigm~\cite{inflation} is a particularly elegant and simple theoretical idea, which explains well many observed features of the observable Universe, such as large scale homogeneity and isotropy, as well as the growth of structure originating from the quantum
fluctuations of the (scalar) inflaton field that drives inflation. The current observational situation~\cite{wmap5} on the absence of any concrete (or, better, statistically significant) evidence of non-Gaussian spectrum of (scalar) perturbations is in agreement with the predictions of the simplest of the inflationary models, that involve a single inflaton field. Admittedly, this situation may change in the future, in case there will be statistically significant evidence on the existence of deviations from Gaussianity. Even if such scenario is realized~\cite{nongauss}, however, it will not mean that the inflationary idea is incorrect, as some times is stated. Most likely, it will mean that the single scalar inflaton model should be modified by including additional ingredients, e.g. other fields,
whose quantum fluctuations could lead to observable and statistically significant deviations from the Gaussian case. Indeed, there are theoretical models involving two or more scalar fields~\cite{twoscalars}, as well as string theory models, e.g. the ekpyrotic scenario~\cite{ekpyrotic} or string gas models~\cite{gas},  which are capable of predicting such deviations.

Since inflation pertains to the early stages of the Universe, it might be either fundamental or effective~\cite{weinberg}, in
the sense of itself being derived from a microscopic theory of quantum gravity as some sort of low-energy approximation. This is clearly the case of string theory, should the latter be assumed to describe correctly the quantum structure of space-time at small scales.
Given that in traditional string theories, the dimensions of space time are higher than four, any inflationary mechanism necessitates consistent compactification schemes, and in this sense it is bound not to be as simple as the original idea. Indeed, there are
many scalar fields in string theory associated with the extra dimensions, the moduli~\cite{stringinfl}, and in this sense their quantum fluctuations in the early Universe might lead to all sorts of complications or even deviations from Gaussianity. Moreover, there are
models~\cite{calabiyauthroats} in string theory whereby inflation is achieved by geometrically and physically involved mechanisms, such as the rolling of fields from the string gravitational multiplet,
down the throats of Calabi-Yau manifolds connecting brane Universes.

In our opinion, there might be simpler mechanisms of inflation which could avoid the compactification problem altogether and thus gain in simplicity. The purpose of this work is to discuss the possibility of one such case, namely that of closed (or open) bosonic strings in first quantized framework ($\sigma$-models), propagating in graviton, dilaton and tachyon backgrounds. Whether such models can be embedded in a phenomenologically realistic framework, from the point of view of particle physics predictions, shall not concern us in this work. What we shall demonstrate, however, is the existence of some very simple conditions that can guarantee inflationary solutions for the target space time, which are re-summed in the Regge slope $\alpha '$.

At this point we remind the reader that the target-space dynamics of strings in a first-quantized framework is based on equations of motion, involving backgrounds of the massless degrees of freedom of the string, which are generated by the vanishing of the Weyl-anomaly coefficients (beta functions)
$\beta^i$ of the world sheet non-linear sigma model, as required by local conformal invariance~\cite{string}. The usual approach consists of seeking perturbative solutions of the vanishing of these beta functions, truncated at most to second order in $\alpha^{'}$. Although such an approach may suffice for a discussion of low-energy (field-theoretic) effects of strings,  all-orders-in-$\alpha '$  re-summation schemes may be essential, in particular when discussing effects at the early Universe, including the inflationary epoch of interest to us here, or in cases where the string mass scale is sufficiently low, e.g. of the order of a few TeV, as is the case in some of the modern approaches to string theories.

To this end, an  approach to obtaining re-summed in $\alpha '$ Weyl anomaly coefficients was proposed
in \cite{AEM1}, in the case of a time-dependent configuration of the bosonic string in metric and dilaton backgrounds. This approach, which is based on a novel functional method on the world-sheet of the string,
leads to homogeneous $\beta^i$: for each of these, all orders in $\alpha^{'}$ are the same power law in time. This suffices to secure conformal invariance as follows: From two-loop (in $\sigma$-model) and higher, the expressions for the beta functions are not unique, but can be modified by a set of
field redefinitions, i.e. a string reparametrization \cite{metsaev} which does not affect the physical
predictions of the theory. As a consequence, if all orders in $\alpha^{'}$ of $\beta^i$ are homogeneous in time
(same power law $(X^0)^{a_i}$), each beta function can be written in the form
\be\label{hombeta}
\beta^i=\sum_{n=0}^\infty\xi_n^i(X^0)^{a_i}(\alpha^{'})^n=A^i(X^0)^{a_i},
\ee
where the constant $A^i$ depends on
the string parametrization, and can be set to zero by choosing specific field redefinitions~\footnote{We note at this stage that homogeneous $\beta$-functions have also appeared in the literature~\cite{mich}, dealing with $\sigma$-models in Anti-de-Sitter target spaces. However, these models do not contain cosmological tachyons or dilatons, and hence they are different in context from the situation discussed here. The interested reader might find a comparison of our formal approach with those cases in \cite{AEM1}.}. This cancellation was
explicitly shown at two loops in \cite{AEM1}, and arguments have been given that it can be performed to all orders. It was found in \cite{AEM1} that the effect of this string reparametrization is to rescale the metric by a constant factor and add another constant to the dilaton, and hence it does not change the time dependence of the configuration. We note that a string reparametrization, generated by field redefinitions, corresponds to changing the renormalization scheme for the calculation of the renormalized energy momentum tensor of the world sheet sigma model~\cite{metsaev,tseytlin}.

The target space corresponding to the configuration found in \cite{AEM1} is a power law expanding Universe, whose dimension $D$ is \emph{not} restricted by \emph{any constraint}, since the above-mentioned $\alpha'$-re-summed
conformal invariance does not involve the vanishing of the tree-level conformal anomaly $(D-26)/6$.
This is a consequence of the fact that the pertinent ($\alpha '$-non perturbative) background configurations involve non-trivial dilatons.
The power law
scale factor becomes a constant, leading to a Minkowski Universe, for a specific choice of the dilaton
amplitude, depending on $D$. An extension of this work, including the antisymmetric tensor \cite{AMT1}, led to the study of optical activity generated by the axion, when the antisymmetric tensor field strength is coupled to an Abelian electromagnetic field.
In this approach, therefore, the \emph{target-space dimensionality} can be chosen to be \emph{four}. If, therefore, an inflationary scenario can be obtained in this framework, in a way consistent with conformal invariance of the pertinent $\sigma$-model,
it will be free from the above-mentioned complications associated with compactification.

In the present work we shall argue that such a possibility can be realized within the above-mentioned $\alpha '$-resummed framework, upon including a tachyon background along side with those of dilaton and graviton.
For the sake of simplicity we shall not consider antisymmetric tensor backgrounds, which do not add anything to the main
issues of interest here, namely the derivation of inflationary metrics.
It should be mentioned that tachyon cosmology has attracted recently a lot of attention, following the works of Sen~\cite{sen} on the r\^ole of tachyons in providing cosmological instabilities in brane world scenarios.

Most of the works on tachyon cosmology involve open-string tachyons, which pertains to the case of brane cosmologies~\cite{sen}, given that the open strings with their ends attached to the brane worlds constitute stringy excitations on the latter, which in broken target-space supersymmetries
can lead to (cosmological) instabilities, represented by open-string tachyons~\cite{type0}. Such instabilities can decay
quickly, so that they are absent in late-times cosmologies.

For our purposes in this work we shall consider closed string tachyons, $T$, which is the simplest model where inflation can be realized.
As we shall demonstrate, the inflationary epoch characterizes target space configurations in which the dilaton $\phi$ and the tachyon $T$ configurations are of similar form, as functions of the cosmic time $X^0$, a sort of ``(anti)alignment'' in theory space of strings, in the sense of a constraint
\begin{equation}
2\phi = -T~.
\label{antialign}
\end{equation}
This suffices to guarantee an inflationary de-Sitter type metric in target space. Deviations from (\ref{antialign}), caused by perturbations of the dilaton and tachyon fields, destroy inflation and lead to power-law Universes. Although the exit and reheating phase
is still an open issue in this approach, nonetheless we believe that this particularly simple model of inflation might lead to further insights on early universe cosmology, within this $alpha'$-resummed framework.

The structure of the article is as follows: in section \ref{sec:2} we discuss the conformal invariant solutions for tachyon, dilaton and graviton backgrounds, within the $\alpha'$-exact framework developed in \cite{AEM1}. In section \ref{sec:3} we discuss first issues related to (well-known) ambiguities that characterize the target-space dynamics in the presence of tachyon backgrounds, namely field redefinition ambiguities in the (low-energy) string effective actions, which leave the perturbative scattering amplitudes unaffected. Then we proceed to describe
the inflationary scenario in our context. We discuss two inflationary solutions, whose physics may be entirely different, associated with their exit phase. One concerns the so-called ``eternal inflation'', which involves a simple de-Sitter exponential metric (of cosmological constant type).
In this case, exit from inflation can be achieved only when the solution is perturbed away from the (anti)alignment configuration (\ref{antialign}), which can be achieved by a (yet unknown) mechanism of either generating masses for the dilaton field, whose decay in standard model particles can cause reheating and exit from the de Sitter phase or through some other type of (stringy) phase transition. The de Sitter phase \emph{per se}, viewed as a classical solution with ``eternal inflation and acceleration'',  does not admit well defined scattering amplitudes, due to cosmic horizons. Hence it is not consistent with perturbative string amplitudes.
In the second case, the inflationary phase is a relatively short epoch interpolating between two flat Minkowski universes. This solution corresponds to a well-defined Scattering matrix and hence it can be consistently accommodated in a perturbative string framework. It does not require the condition (\ref{antialign}) but a much softer version of it, namely a condition relating
only the amplitudes of the corresponding fields. In this case, the exit from inflation occurs already at the level of a classical solution, but particle (or rather string) production can be discussed in a rather standard way~\cite{gubser}, involving Bogolubov coefficients of the non-trivial ``in'' and ``out'' vacua.
Conclusions and Outlook are presented in section \ref{sec:4}. Finally, some technical aspects of our work, such as a description of the functional (world-sheet) method used, and the details of some field redefinitions that guarantee conformal invariance of the solutions, as well as passage from String to Einstein frames are presented in three Appendices A, B, C respectively.

\section{$\alpha^{'}$-resummed conformal invariance of lowest mass-level string states\label{sec:2}}

In this section we shall discuss the theoretical framework for
obtaining exact (in the Regge slope $\alpha '$) solutions to the conformal invariance conditions of a stringy
bosonic $\sigma$-model in non-trivial tachyon, dilaton and metric backgrounds.
Caution should be exercised here with precise meaning of the term re-summation in $\alpha '$-corrections.
This does not mean that these are truly non-perturbative string solutions.
In the present work and in \cite{AEM1}, we have ignored all massive string states, with masses that are
integer multiples of the string mass scale $M_s=1/\sqrt{\alpha '}$. In our approach to string cosmology, we consider only an effective field theory level, in the sense of keeping only excitations with momenta
small as compared to the string scale. These are assumed to be the relevant degrees of freedom
for inflation. Indeed, as we shall discuss below (c.f. (\ref{hubbleinfl})), the typical Hubble scale during inflation in our model can be a few orders of magnitude smaller than the string mass scale, and we assume that only modes with energies and momenta lower than this scale are relevant in our discussion. In such a framework it makes sense to consider the effective field theory limit of strings by ignoring the massive string states, for which we make the assumption that they do not play an important r\^ole in inflation. But we have to stress that this is only an assumption. In the string Very Early Universe Massive string states could indeed play a r\^ole, which however goes beyond the scope of the present work.

However, the above assumption does not preclude the necessity for going beyond the truncation to a given order in an $\alpha '$-expansion of the Weyl anomaly coefficients for the low-lying-mass level degrees of freedom (tachyons, dilatons and gravitons here). This is because, for cosmological backgrounds
depending only on time, the various orders in the
$\alpha '$-expansion yield terms independent of $\alpha '$, and hence they are all of the same order
(``homogeneity'' property (\ref{hombeta})). In this sense, one should seek for a method of computing the $\beta$-functions of the stringy $\sigma$-model in a re-summed fashion, in similar spirit to the re-summation of the leading $1/N$ expansion in gauge theories with ``large'' number of fermions. This is what we do in \cite{AEM1} and  here.

The method is based on a novel functional method for discussing running of physical quantities with certain parameters of a
quantum theory, in which the ultraviolet cutoff is fixed. Hence the method is different from the Wilsonian renormalization group approach, although in certain circumstances contact with this approach can be made. This
method was first developed in \cite{polonyi}
for four-dimensional field theories, and subsequently applied consistently to stringy two-dimensional $\sigma$-models~\cite{AEM1,liouville},
which is the formalism we follow in the present work.

\subsection{World sheet quantum theory}\label{sec:fm}

We commence our analysis with the following time-dependent world sheet bare sigma-model in metric, dilaton and quadratic tachyon backgrounds
\bea
S_\sigma&=&\frac{1}{4\pi\alpha^{'}}\olo \xi \sqrt\gamma\Big\{\gamma^{ab} g_{\mu\nu}(X^0) \partial_a X^{\mu}\partial_b X^{\nu} \nn
&&~~~~~~~~~~~+4\pi\alpha^{'}\lambda\Lambda^2 \left(X^0\right)^2+\alpha^{'}R^{(2)}\phi(X^0)\Big\},
\eea
where $\lambda$ is our running parameter, which controls the amplitude of the tachyon mass, proportional
to the {\it fixed} world sheet cut off $\Lambda$. Our approach
consists~\cite{AEM1} in finding the evolution of the quantum theory with $\lambda$, and to look for a solution which is
$\lambda$-independent. We will see that the corresponding configuration has conformal properties  in a resummed $\alpha'$-framework.
In the absence of tachyons, the $\alpha '$-resummed  configuration for graviton and dilatons
was first found in \cite{AEM1},  and has the form
\bea\label{nonperturb1}
g_{\mu\nu}&=&\frac{A}{(X^0)^2}\eta_{\mu\nu}\nn
\phi&=&\phi_0\ln\left(\frac{X^0}{\sqrt{\alpha'}}\right) .
\eea
where the constant $A$ has dimensions of [Length]$^2$, and hence can be expressed in units of $\alpha '$.
The latter backgrounds lead to Weyl-invariance beta functions homogeneous in $X^0$, to all orders in $\alpha^{'}$,
and as explained in \cite{AEM1}, this was sufficient to satisfy conformal invariance.
In the present work, we consider the $\alpha'$-resummed configuration (\ref{nonperturb1}) for the metric and dilaton,
and look for the corresponding tachyon background which will respect the homogeneity of beta function and lead to Weyl invariant configurations, in the sense of demonstrating that there is sufficient freedom to redefine the backgrounds so as to
achieve one renormalization scheme~\cite{metsaev,tseytlin}, in which the pertinent Weyl anomaly coefficients of the tachyon, dilaton and target-space metric backgrounds vanish.

We define in Appendix A the proper graphs generator functional $\Gamma$ of the theory, representing the dressed
theory, which includes quantum corrections. We show in the same Appendix the derivation of the exact evolution equation
of $\Gamma$ with $\lambda$, and find for a flat world sheet
\be\label{evolG}
\dot\Gamma=\Lambda^2\int d^2\xi (X^0)^2+\Lambda^2
\mbox{Tr} \left\{\left(\frac{\delta^2 \Gamma}{\delta X^0 \delta X^0}\right)^{-1}\right\}
\ee
where a dot denotes a derivative with respect to $\lambda$.
We stress here that, although this self-consistent equation technically looks like a
Wilsonian exact evolution equation, it is
actually very different in essence. Indeed, our approach makes use of the {\it fixed} world sheet cut off $\Lambda$, and
our running parameter is $\lambda$. As a consequence, we describe here a kind of differential Schwinger-Dyson equation,
which avoids the use of a non-physical world sheet cut off.

The next step is to assume, in the framework of the gradient expansion, a functional dependence for $\Gamma$. We consider
the following form
\be
\Gamma[X]=\frac{1}{4\pi\alpha'} \olo \xi \left\{\delta^{ab}
\frac{A}{(X^0)^2}\eta_{\mu\nu}\partial_a X^\mu \partial_b X^\nu+4\pi\alpha'\la T(X^0)\right\},
\ee
and insert the latter expression into the evolution equation (\ref{evolG}). We thus obtain
\be
\dot T(X^0)= (X^0)^2 + \alpha'\frac{(X^0)^2}{2A} \ln \left[1+\frac{A}{2\pi\alpha'(X^0)^2 T''(X^0) }\right],
\ee
where a prime denotes a derivative with respect to $X^0$.
Note that no divergence is present here, as the fixed world sheet cut off $\Lambda$ appears in the definition of the tachyon.
One can see that a $\lambda$-independent solution, satisfying $\dot T=0$, is:
\be \label{tachconf}
T= \frac{1}{2\pi\alpha'}\frac{A}{1-e^{-2A/\alpha'}}\ln\left(\frac{X^0}{\sqrt{\alpha'}}\right)
\equiv \tau_0\ln\left(\frac{X^0}{\sqrt{\alpha'}}\right),
\ee
For $A > 0$ we have that $\tau_0 > 0$.

We discuss the corresponding conformal (Weyl-invariance) properties
of the configuration (\ref{nonperturb1},\ref{tachconf})
in the next subsection.

\subsection{Weyl invariance}\label{sec:wi}

In this subsection we shall study the conformal properties of the following configuration
\bea\label{config}
g_{\mu\nu}&=&\frac{A}{(X^0)^2}\eta_{\mu\nu}\nn
\phi&=&\phi_0\ln\left(\frac{X^0}{\sqrt{\alpha'}}\right)\nn
T&=&\tau_0\ln\left(\frac{X^0}{\sqrt{\alpha'}}\right).
\eea
and provide arguments that, upon suitable redefinitions of the background fields, one
can always find a renormalization group scheme in which the Weyl anomaly coefficients vanish at arbitrary order in
$\alpha '$-expansion.\\
At one-loop, the Weyl anomaly coefficients (beta functions) are,
\bea\label{1loop}
\beta^g_{\mu\nu}&=&\alpha'R_{\mu\nu}+2\alpha'\nabla_\mu\nabla_\nu\phi
-\alpha^{'}\partial_\mu T\partial_\nu T+{\cal O}(\alpha')^2\nn
\beta^\phi&=&\frac{D-26}{6}-\frac{\alpha'}{2}\nabla^2\phi+\alpha'\partial^\mu\phi\partial_\mu\phi+{\cal O}(\alpha')^2\nn
\beta^T&=&-2T-\frac{\alpha'}{2}\nabla^2 T+\alpha'\partial^\mu\phi\partial_\mu T+{\cal O}(\alpha')^2,
\eea
and one can easily check that, besides the tree-level term $-2T$ in $\beta^T$, the configuration (\ref{config}) leads
to homogeneous beta functions:
\bea\label{betas}
\beta^g_{00}&=&-\frac{\alpha^{'}}{(X^0)^2}(D-1+\tau_0^2)+{\cal O}(\alpha^{'})^2\nn
\beta^g_{ij}&=&\frac{\alpha^{'}\delta_{ij}}{(X^0)^2}(D-1+2\phi_0)+{\cal O}(\alpha^{'})^2\nn
\beta^\phi&=&\frac{D-26}{6}+\frac{\alpha^{'}}{2}\left(D-1+2\phi_0\right)\frac{\phi_0}{A}+{\cal O}(\alpha^{'})^2\nn
\beta^T&=&-2T+\frac{\alpha^{'}}{2}\left(D-1+2\phi_0\right)\frac{\tau_0}{A}+{\cal O}(\alpha^{'})^2
\eea
Higher orders in $\alpha^{'}$ are also homogeneous, as can be readily seen by power counting: whatever power of the
Ricci or Riemann tensor one considers, multiplied by powers of the covariant derivatives of the tachyon or the dilaton,
the contraction of the indices with the metric or its inverse will always lead to the same power of $X^0$.\\
Once the $\beta$-functions of the theory are homogeneous, one can rather easily make them vanish, based on the fact that their definition is not unique: One is always free to make general field redefinitions, under which the physics of the theory is invariant; this leads to a different definition of the $\beta$-functions. In our case we employ a field redefinition, that can also cancel the inhomogeneous linear term in $\beta^T$:
\bea \label{redef}
g_{\mu\nu} &\rightarrow& \tilde{g}_{\mu\nu} = g_{\mu\nu} + \ap g_{\mu\nu}
\Big(a_1 R + a_2 \partial \phi \cdot \partial \phi + a_3 \partial \phi \cdot \partial T\nn
&&~~~~~~~~~~~~~~~~~~~~~~~~+a_4 \partial T\cdot \partial T + a_5 \nabla^2\phi +a_6\nabla^2 T\Big)  \nn
\phi &\rightarrow& \tilde{\phi} = \phi + \ap
\Big(b_1 R + b_2 \partial \phi \cdot \partial \phi + b_3 \partial \phi \cdot \partial T\nn
&&~~~~~~~~~~~~~~~~~~~~+b_4 \partial T\cdot \partial T + b_5 \nabla^2\phi +b_6\nabla^2 T\Big)  \nn
T &\rightarrow& \tilde{T} = T+\ap a R T+\ap
\Big(c_1 R + c_2 \partial \phi \cdot \partial \phi + c_3 \partial \phi \cdot \partial T\nn
&&~~~~~~~~~~~~~~~~~~~~~~~~~~+c_4 \partial T\cdot \partial T + c_5 \nabla^2\phi +c_6\nabla^2 T\Big),
\eea
\nin where $a$, $a_1$,..., $a_6$, $b_1$,..., $b_6$ and $c_1$,... $c_6$ are constants to be determined. For our configuration, these field redefinitions do not change the $X^0$-dependence of the fields, since
\bea
\tilde{g}_{\mu\nu} &=& d_1g_{\mu\nu}  \nn
\tilde{\phi} &=& \phi + d_2 \nn
\tilde{T} &=& d_3 T+d_4,
\eea
where $d_1,d_2,d_3,d_4$ are constants.
The details of the effects of this field redefinition on the $\beta$-functions are given in Appendix B. As explained there, the term $\ap a R T$ that appears in the above redefinition of $T$ plays a different role than the rest of the terms in the field redefinitions: it cancels the inhomogeneous term in $\beta^T$. In more detail, it causes two changes in the $\beta$-functions: it produces a new term in $\beta^{g}_{00}$ that is homogeneous to the rest of the terms in it (i.e. is proportional to $(X^0)^{-2}$) and a new term in $\beta^T$ that is linear to $T$ and can thus cancel the inhomogeneous term, $-2T$, as long as $a$ is chosen to be:
\be \label{a}
a=\frac{2A^2}{(\ap)^2} \frac{1}{ (D-1) \left[D-D^2 -4(D-1)\phi_0 +(D-2)\tau_{0}^{2}\right]},
\ee
The rest of the effect of this redefinition is to add new terms to the $\beta$-functions that are homogeneous to the already existing terms.
Taking into account the initial two-loop $\beta$-functions, which have the same homogeneity, in the end we have new $\beta$-functions with the structure (see Appendix B for more details):
\bea
\tilde\beta^{g}_{00} &=&\frac{\tilde E_1}{(X^0)^2}+{\cal O} \left({\ap}^3\right) \nn
\tilde\beta^{g}_{ij} &=&\frac{\tilde E_2}{(X^0)^2}\delta_{ij}+{\cal O} \left({\ap}^3\right) \nn
\tilde\beta^\phi &=&\tilde E_3 +{\cal O} \left({\ap}^3\right) \nn
\tilde\beta^T &=&\tilde E_4 +{\cal O} \left({\ap}^3\right),
\eea
where the constants $\tilde E_1, \dots \tilde E_4$ are linear combinations of $(a_i,b_i,c_i)$. It is then always possible to choose the 18 constants $(a_1,\dots,a_6,b_1,\cdots,b_6,c_1,\cdots,c_6)$ in order to cancel the $\tilde{E}_i=0$, and thus make the $\beta$-functions vanish.
This analysis can be repeated to higher orders in $\ap$ \cite{tseytlin}, thus proving that our configuration, which is valid to
all orders in $\ap$, respects conformal invariance to all orders in $\ap$.

 Some remarks are in order at this juncture, concerning the extension of the above ideas to open strings, which are fashionable recently in view of the brane world cosmological scenarios.
In such cases, the tachyon excitations describe instabilities of the brane vacuum, and correspond to open string tachyonic excitations, with the ends of the string being attached to the brane world.
From an open-string $\sigma$-model point of view the extension of the above conclusions, namely the
form of the tachyon field is straightforward. The functional method presented above does not require any specific assumptions other than the possibility of working with flat world sheet geometries, and in this sense the fixed point solutions discussed above are valid, provided of course the relevant $\beta$-functions are \emph{homogeneous}. Thus, although technically the various terms in the tachyon $\beta$-functions are different from the corresponding terms in the closed string case at various orders in the $\alpha '$-expansion, nevertheless the homogeneity requirement guarantees the existence of an appropriate renormalization scheme (non perturbatively in $\alpha '$) in which the solution for the tachyon (\ref{tachconf}) satisfies conformal invariance to all orders in $\alpha '$.

In the open string case, the inflationary scenario can also be achieved upon the requirement of the anti-alignment condition (\ref{antialign}) between the open tachyon and the dilaton fields.
However, the reader should bear in mind that from a $\sigma$-model view point the dilaton, as a field of the gravitational multiplet of the string, pertains to the bulk of the world sheet, whilst the open string tachyon is associated with a world-sheet boundary operator. But such technical details do not alter the main conclusions on the conditions for target-space inflation, which carry through intact from the closed string case. In what follows therefore, we shall concentrate for concreteness to closed Bosonic strings, but we shall inject from time to time a comparative discussion on the open-string tachyon case.

\section{Space time effective-field-theory action and inflation \label{sec:3}}

\subsection{Field Redefinition Ambiguities and closed-string Tachyon backgrounds}

Starting from our $\alpha'$-resummed configuration (\ref{config}),
we redefine the time coordinate as $X^0=\sqrt{\alpha'}\exp(-X'^0/\sqrt{\alpha'})$.
The corresponding
Jacobian arising in the partition function (defined in Appendix A) is then responsible for
an additional linear dilaton $\tilde\phi_0 X'^0/\sqrt{\alpha'}$,
which, together with $\phi_0\ln(X^0/\sqrt{\alpha'})=-\phi_0 X'^0/\sqrt{\alpha'}$, leads to the linear dilaton
$-\phi_1 X'^0/\sqrt{\alpha'}$, where $\phi_1=\phi_0-\tilde\phi_0$.
In terms of the new time coordinate, and after rescaling all the coordinates by $\sqrt{\ap/A}$, $
\left(X'^0,X^i\right) ~\rightarrow~ \left(\tilde{X}^0,\tilde{X}^i\right) ~=~ \left(\sqrt{\frac{A}{\ap}} X'^0,\sqrt{\frac{A}{\ap}}X^i\right)~$,
so that the time component of the metric is normalized to 1, as in a standard Robertson-Walker Universe,
the relevant configuration becomes:
\bea\label{configbis}
g_{00}&=& \eta_{00}\nn
g_{ij}&=& \exp\left( \frac{2\tilde{X}^0}{\sqrt{A}}\right) ~\eta_{ij}\nn
\phi&=&-\phi_1 \tilde{X}^0/\sqrt A\nn
T&=&-\tau_1 \tilde{X}^0/\sqrt A,
\eea
and the $\sigma$-model action reads:
\begin{eqnarray}
S_\sigma&=&\frac{1}{4\pi \alpha '}\int d^2\sigma\sqrt{\gamma}
\Big(\eta_{00}~\partial_a \tilde{X}^0 {\partial}^a \tilde{X}^0  + \eta_{ij}~e^{\frac{2\tilde{X}^0}{\sqrt{A}}}
\partial_a \tilde X^i \partial^a \tilde X^j\nonumber \\
&& ~~~~~~~~~~~~~~~~~~~~-\alpha ' \phi_1  R^{(2)}\frac{\tilde{X}^0}{\sqrt A}
-4\pi\alpha'\Lambda^2\tau_1\frac{\tilde{X}^0}{\sqrt A} \Big)~.
\label{sigmamodel2}
\end{eqnarray}
The configuration (\ref{configbis}) represents a $\sigma$-model-frame de Sitter (inflationary) metric  with Hubble parameter $H_I =1/\sqrt{A}$. \footnote{The $\sigma$-model-frame metric $g_{\mu\nu}\sim (X^0)^{-2}\eta_{\mu\nu}$ is de Sitter without the need for the coordinate redefinition $X^\mu\rightarrow\tilde X^\mu$ (one recognizes $X^0$ as the conformal time coordinate). The redefinition $X^\mu\rightarrow\tilde X^\mu$ is useful as the dilaton field becomes linear in these coordinates, and the metric takes its standard FRW form at the same time.}

For reasons that will become clear soon, we need an explicit knowledge of the lowest non-trivial order (in an $\alpha '$, derivative expansion) of the effective target-space action in the presence of (closed string) tachyon backgrounds. We remark at this point that, although for open-string rolling tachyons attached to Dp-brane models~\cite{sen} there has been significant progress~\cite{tacheff} in understanding the emergence of (the so-called Dirac-Born-Infeld (DBI) type) actions, which in flat brane space-times acquire the closed form
\begin{equation}
S_T^{\rm open} = \int d^px U(T)\sqrt{1 + \partial_\mu T \partial^\mu T}~,
\label{opentach}
\end{equation}
with $U(T)$ of appropriate form,
this is not the case for closed-string tachyon effective actions,
where the situation is far from being conclusive, even to lowest non-trivial order in $\alpha'$ \cite{tseytlin,swanson}.

In a derivative expansion, the lowest non-trivial order
of the target space-time effective action describing the dynamics of the closed string
dilaton $\phi$, graviton $g_{\mu\nu}$ and tachyon $T$ backgrounds
can be expressed generically as:
\bea\label{Sigma}
S&=&\int d^D x\sqrt{-g}e^{-2\phi}\Big\{\frac{D-26}{6\ap} + f_0(T)
+f_1(T)R+4f_2(T)\partial_\mu\phi\partial^\mu\phi\nn
&&-f_3(T)\partial_\mu T\partial^\mu T-f_4(T)\partial_\mu T\partial^\mu\phi\Big\},
\eea
where $d^Dx=dx^0 d{\bf x}$ ($x^\mu$ denotes the zero mode of $X^\mu$),
and the functions $f_i (T)$ depend on the tachyon but not its derivatives.

The reader should notice that the (on-shell) string-scattering amplitudes cannot provide sufficient information to determine the form of $f_i(T)$, due to the possibility of field redefinitions that leave the (perturbative) string scattering matrix invariant. Thus one has to find alternative arguments in order to determine these functions. There are various ways to do this~\cite{tseytlin,swanson}, including the postulate of symmetries, such as the invariance of the D-dimensional string effective action by $O(D-1,D-1)$ transformations~\cite{meissner}, where $D-1$ is the number of spatial dimensions, but the situation seems not conclusive.

Depending on the form of $f_i(T)$ one can obtain different solutions of the equations of motion
of the string effective actions that are equivalent to the Weyl invariance conditions in the sense
of local field redefinitions that leave the (perturbative) string scattering amplitudes unaffected. Since the latter carry physical information, the class of the allowed solutions that are connected by field redefinitions must be such that the scattering matrix (S-matrix) is well defined. This requirement excludes the case of space times with horizons from being obtained from such perturbative string theory constructions~\cite{banks}. However, when one considers non perturbative string theory solutions,
such arguments do not apply. In this sense, the issue of a de Sitter solution, representing a cosmological constant or, equivalently ``eternal'' inflation solution, in string theory cannot be answered by means of string scattering amplitudes. The latter are not well defined in such a case  due to the existence of a cosmological horizon that prevents the definition of asymptotic states~\cite{challenge}.

However, it is possible that one starts from perturbative string amplitudes and arrives at a form of a target-space effective action which, among its solutions, also admits de Sitter inflationary solutions with space time horizons, which do not have well defined S-matrix, and thus carries more information than the one contained in perturbative string theory.
In this work, in the next subsection, we shall first discuss a case of eternal inflation in our tachyon-dilaton scenario, upon the
anti-alignment condition (\ref{antialign}) in solution space.
However, as we shall argue in the next section the inflationary epoch can indeed be considered as a brief epoch in an interpolating solution
between two asymptotically flat regions of space time, a case which can characterize perturbative string
theory as well, since the S-matrix is well defined. Such a case corresponds to a
different class of functions $f_1(T)$. It is this solution that we shall accept as a self-consistent solution in our approach.

We now remark that in ref.~\cite{tseytlin},
the simple choice $f_1(T)=e^{-T}$ is made, based on the definition of the target-space effective action
in the $\sigma$-model frame as the derivative of the
world-sheet partition function with respect to the (logarithm of the) ultraviolet cutoff $\ln \epsilon$,
with
\be
\epsilon^{-2} = \Lambda^2{\cal A}
\ee
where $\Lambda$ is a covariant (momentum space, UV) cutoff on the world-sheet and
${\cal A}$ is the world sheet surface area. We have then
\begin{equation}\label{effact}
S = - \left[ \frac{\partial Z}{\partial \ln \epsilon}\right]_{\epsilon=1} = {\beta}^i \cdot \frac{\delta Z}{\delta g^i}
\end{equation}
where $Z$ is the world-sheet partition function of the $\sigma$-model in $\phi$, $g_{\mu\nu}$ and $T$ backgrounds:
\bea
Z[g_{\mu\nu},\phi,T]
=\int [dX] \exp\left\lbrace \frac{1}{4\pi\alpha'}\int d^2 \xi \sqrt{\gamma}
\left[  \gamma^{ab} g_{\mu\nu}\partial_a X^\mu \partial_b X^\nu + \ap R^{(2)} \phi+
4\pi\alpha'\Lambda^2 T \right]\right\rbrace
\eea
Pulling out the world-sheet zero modes of $\phi$ and $T$ we can write
\be
Z[g,\phi,T,\Lambda] \sim \int [dx] \sqrt{-g} e^{-2\phi-\epsilon^{-2}T}e^{-W}
\ee
where the functional is defined using the background field method~\cite{tseytlin}.
Note that all the fields in $Z$, as well as $Z$ itself, depend on the cutoff, $\Lambda$. In our approach~\cite{polonyi,AEM1}
we keep the \emph{cutoff fixed}.\\
To lowest order in an $\alpha '$ expansion, one can show that
\bea\label{ea2}
W&=& \frac{1}{2} \gamma \ln \epsilon + {\cal O} \left(\ln^2\epsilon\right) \nn
\gamma &=& \frac{2}{3}(D-26) - \ap \epsilon^{-2}\nabla^2 T -2\ap \nabla^2 \phi-\ap R
+ {\cal O} \left(\ap^2\right)
\eea
which, upon appropriately absorbing the fixed (in our approach) cutoff factors $A \Lambda^2  $ into $T$,
would lead, by virtue of (\ref{effact}),  to an effective action of the form :
\be\label{effact2}
\tilde S = \int d^Dx\sqrt{-g} e^{-2\phi-T} \left[\frac{D-26}{3} -2T - \frac{\ap}{2} \nabla^2 T - \ap \nabla^2 \phi - \miso \ap R + {\cal O} (\ap^2)  \right]
\ee
It can be readily seen that this is of the
form (\ref{Sigma}),
and by comparison we deduce that in this case
\begin{equation}
f_1(T)=e^{-T}~.
\label{f1}
\end{equation}
In fact, one can even give a closed formal expression for the effective action in terms of the $\beta$-functions
 for the background fields, to arbitrary order in $\alpha '$~\cite{tseytlin}:
\be\label{sbeta}
S = \int d^Dx \sqrt{-g}e^{-2\phi - T}\left(\beta^T + 2\beta^\phi -\frac{1}{2}g^{\mu\nu}\beta^g_{\mu\nu}\right)~.
\ee
There are two important issues, though, which forced Tseytlin~\cite{tseytlin}  to dismiss this construction.
The first is that the tachyon kinetic terms in (\ref{sbeta}), even after the passage to the Einstein frame (c.f. Appendix C),
appear with the wrong sign (ghost-like fields), unlike the dilaton terms which in the Einstein-frame effective action appear with the canonical kinetic terms. The second reason is
that the linear tachyon-dependent terms in (\ref{effact2}), which stem from the corresponding terms in the tachyon $\beta$-function
(\ref{betas}), imply that for $T \ne 0$ the usual perturbative bosonic-string vacuum
($D=26$, $\phi = T = 0$ and $g_{\mu\nu}=\eta_{\mu\nu}$) is not a stationary point of the action (\ref{sbeta}).
This is a well-known problem associated with the non-trivial tachyon tadpoles in this case.
Tseytlin in \cite{tseytlin}, attempted to subtract the tachyon tadpole in a renormalization-group (RG) invariant way but
the analysis was not conclusive in that a solution which resolved both of the aforementioned problems could not reproduce the
Weyl-anomaly coefficient for the tachyon, and thus was not compatible with the (perturbative) on-shell scattering amplitudes.

In our approach, it is precisely the tachyon-tadpole term that creates the unwanted inhomogeneities in the $\beta$-functions. However,
we have not removed it in a RG invariant way, on the contrary we have determined a specific field redefinition for the
tachyon field (\ref{redef},\ref{a}) in which it is absent from the $\beta^T$ function. Our $\alpha'$-resummed approach is based precisely on the homogeneity properties of the $\beta$-functions, and hence it is natural to expect that, in our framework,
the effective action (\ref{sbeta}), constructed (formally) out of these $\beta$-functions, contains the right information.
For us, there is no reason for the perturbative string vacuum to be included in our
$\alpha '$-resummed scheme. In fact, in the way we have defined our ground state of the string, based on the
novel functional method of \cite{AEM1,polonyi}, the resulting string configurations are valid for
dimensions of the target space time other than the critical value 26.
We have even argued in \cite{AEM2} that our solutions may not even be smoothly connected to the trivial vacuum
because the \emph{spectrum} of the relevant target-space \emph{dimensionality} $D$ can be \emph{discrete}.
The above arguments support the idea that our vacuum configuration for the bosonic string
does not include the trivial vacuum $\phi = 0,~ T=0$ and thus it probably
constitutes an entirely different sector of (\emph{non-critical}) string theories, with the time $X^0$ playing the r\^ole
of the Liouville mode. This is evident from (\ref{sigmamodel2}).

With this in mind, we can therefore assume that the choice (\ref{f1}) is valid within our $\alpha '$-resummed framework, but not the action (\ref{sbeta}). This assumption seems necessary in order to avoid unitarity problems in the tachyon sector, in the sense of the absence of ghost-like fields,  which are present if the form (\ref{sbeta}) is used~\cite{tseytlin}, as mentioned above.

To determine the ghost-free structure of the effective action in the tachyon-dilaton sector
we first note that, because the metric redefinition (\ref{re}) in Appendix C, which allows the passage to the Einstein frame, involves the tachyon, there are induced kinetic terms for the latter. In the $\sigma$-model frame effective action (\ref{Sigma}), the dilaton appears with the ``wrong'' sign ($f_2(0)=1$), whereas the sign of the tachyon kinetic term depends on the function $f_3$. The effective action in the Einstein-frame obtained from (\ref{Sigma}) reads:

\bea\label{einstnd}
S^E=\int d^D x\sqrt{-g} \Bigg\{ R  + e^{\frac{4\phi}{D-2}} \left(f_1\right)^{\frac{-D}{D-2}} \left[\frac{D-26}{6\ap} + f_0(T)\right]-\left[\frac{4(D-1)}{D-2} - \frac{4f_2}{f_1} \right]\partial \phi \cdot \partial \phi &&\nn
 - \left[\frac{D-1}{D-2}\left(\frac{f_1'}{f_1}\right)^2+\frac{f_3}{f_1} \right]\partial T \cdot \partial T  + \left[\frac{4(D-1)}{D-2}\frac{f_1'}{f_1} -\frac{f_4}{f_1}\right] \partial\phi\cdot \partial T \Bigg\}~~~&&
\eea

Indeed there is an induced kinetic term for the dilaton field, $-\frac{4(D-1)}{D-2}\partial\phi\cdot\partial\phi$ with the correct sign, but to make sure that the signs are correct for both fields, we need to diagonalize this action. To do this, we redefine the tachyon field as $T~\rightarrow~\tilde T = -\miso \phi + f(T)$ with $f(T)$ a function that satisfies: $f'=\frac{(D-1)(f_1')^2+(D-2)f_1f_3}{4(D-1)f_1f_1'-(D-2)f_1f_4}$. Under this redefinition, the action takes the following form in the Einstein frame:

\bea\label{einst}
&&S^E~=\int d^D x\sqrt{-g} \Bigg\{ R + e^{\frac{4\phi}{D-2}} \left[f_1(T)\right]^{\frac{-D}{D-2}} \left[\frac{D-26}{6\ap} + f_0(T)\right]\nn
&&~~~~-\left[\frac{4(D-1)}{D-2} - \frac{4f_2}{f_1} - \frac{\left[4(D-1)f_1'-(D-2)f_4 \right]^2}{2(D-2)\left[(D-1)(f_1')^2+(D-2)f_1f_3\right]} \right]\partial \phi \cdot \partial \phi \nn
&&~~~~~ -  \frac{\left[\frac{4(D-1)}{D-2}f_1'-f_4\right]^2}{\frac{D-1}{D-2}(f_1')^2+f_1f_3}\partial \tilde T \cdot \partial\tilde T \Bigg\}
\eea

Now the kinetic terms for both fields, $\phi$ and $\tilde T$, have the right signs, as long as the functions $f_i$ satisfy the following relations (for all values of $T$):

\bea \label{restrictions}
&& f_2+\frac{\left[4(D-1)f_1'-(D-2)f_4 \right]^2}{8(D-2)\left[(D-1)\frac{(f_1')^2}{f_1}+(D-2)f_3\right]} ~<~ \frac{D-1}{D-2} f_1 \nn
&& f_3+\frac{D-1}{D-2}\frac{(f_1')^2}{f_1} ~>~0
\eea

\nin The requirement of no ghost fields doesn't place any further restrictions to the generic functions $f_i(T)$ that appear in the sigma-model action (\ref{Sigma}).

It is important to remark at this point that
explicit knowledge of the entire effective action for the tachyon is \emph{not} required in order to discuss our inflationary scenarios.

\subsection{``Eternal Inflation'' or Cosmological-Constant (de Sitter) Vacuum in closed Strings}

To understand the emergence of inflationary solutions from the effective action (\ref{Sigma}),
the reader should first notice that the dynamics of the metric sector of this action
does not have the canonical Einstein-Hilbert form, as a result of the dilaton and tachyon dependent prefactors:
\be\label{sg}
S_g=\int d^D x\sqrt{-g}\exp\{-2\phi+\ln f_1(T)\}R~,
\ee
in the $\sigma$-model frame. In the Einstein frame~\cite{aben} (c.f. Appendix C), on the other hand, which is obtained by performing
appropriate field redefinitions of the metric tensor, the Einstein-Hilbert part of (\ref{Sigma}) assumes its canonical form.
It is in the latter frame that cosmology is discussed~\cite{aben}. These features are important for our analysis. We seek configurations for which the Einstein and $\sigma$-model frame metrics coincide, so that the inflationary solution (\ref{configbis})
characterizes the physical (Einstein-frame) Universe.

In such a case we obtain for the Einstein-Hilbert part (\ref{sg}) of the effective action (\ref{Sigma}) for our configuration (\ref{configbis}):
\be\label{prefact}
S_g=\int d^D x\sqrt{-g}\exp\left((\tau_0+2\phi_0)\ln\frac{x^0}{\sqrt \ap}\right) R,
\ee
Hence, if (c.f. Appendix C for details):
\begin{equation}
\tau_0+2\phi_0=0~,
\label{condition2}
\end{equation}
\emph{i.e}. there is (anti)alignment between dilaton and tachyons in the space of solutions (\ref{antialign}),
then the  Einstein-frame metric~\cite{aben} (in which the Einstein-Hilbert action assumes its canonical form, with no
dilaton and tachyon dependent pre-factors) is the same as the $\sigma$-model-frame metric, and
on account of (\ref{configbis}),
we have \emph{inflation}, with the scale factor varying with the canonical Robertson-Walker cosmic time (c.f. (\ref{ty}) in Appendix C)
as:
\begin{equation}
a (t) = a_0 \,\exp\left( \frac{t}{\sqrt{A}}\right)~.
\label{inflsol}
\end{equation}
The relevant Hubble parameter is therefore:
\be
H_I = \frac{1}{\sqrt{A}}~,
\label{hubbleinfl}
\ee
where we remind the reader that $A$ has dimensions of $\alpha '$.
Thus we observe that the amplitude $A > 0$ of the graviton in the $\sigma$-model-frame configuration (\ref{config})
determines the scale of the Hubble parameter during inflation, and is a free parameter of our model. This is a positive feature for phenomenology. Indeed, as already mentioned, from the WMAP five-year data on CMB temperature fluctuations measurements~\cite{wmap5}, we can deduce that during inflation one must have $H_I < 10^{-5}\, M_{\rm Planck}$, with the Planck mass scale $M_{\rm Planck} \sim 10^{19}$~GeV. Assuming that the string scale is of the same order as the Planck scale, we then have that $A/\alpha' > 10^{10}$.

We also note at this point that a cancellation between tachyon and dilaton effects was already used in \cite{yang_zwiebach}, although there the  configuration satisfied directly one-loop Weyl invariance, and lead to different conclusions on a metric that resulted in a big crunch of the rolling tachyon Universe.

The solution (\ref{configbis},\ref{condition2}) appears to describe \emph{eternal} inflation and acceleration of the Universe, as is
standard in a de Sitter Universe. Such universes are characterized by \emph{cosmic horizons}, and as such the concept of well-defined asymptotic states of a quantum field theory, and hence of an S-matrix, are in trouble. This was presented as a challenge for (perturbative) string theory~\cite{challenge}, which by definition is based on an S-matrix formalism.
In our approach we have started from the perturbative (on-shell) S-matrix approach, and by performing local field redefinitions of the background fields  (order by order in an $\alpha '$-expansion), which leave the perturbative S-matrix elements invariant,  to arrive at a renormalization scheme in which the homogeneous (and non perturbative) Weyl anomaly coefficients vanish.

The constraint (\ref{condition2}) spans a practically \emph{zero density subspace} and constitutes an isolated region in the space of solutions, which thus explains the ill-defined nature of the (perturbative) S-matrix elements in such a ($\alpha'$-resummed) de Sitter case.
We thus see that in this framework we can have inflation in the physical case only if both tachyon and dilaton fields are present, canceling each other (\ref{condition2}), (\ref{antialign}) at a \emph{classical} level in target space.
The eternally accelerating de Sitter solution (\ref{configbis}),(\ref{condition2}) may be viewed as representing a cosmological constant dominating vacuum in string theory.

\subsection{End of inflation and reheating: some speculations }

In principle exit from this classical phase can be achieved through some (yet unknown)
phase-transition mechanism, in which the tachyon decays to zero asymptotically in space time. In such a way the anti-alignment
condition (\ref{condition2}) is disturbed, and the Universe turns into a power-law expanding or contracting one, depending on the
values of the dilaton field~\cite{AEM1}.

Indeed, if one does not impose the constraint (\ref{condition2}) in the space of solutions (\ref{configbis}),  then the
resulting Einstein-frame Universe (obtained~\cite{aben} by a redefinition of the metric tensor such that the exponential pre-factor in front of the Einstein scalar curvature term in (\ref{prefact}) is unity) is always a power-law one, in the sense that its scale factor when expressed in terms of the cosmic Robertson-Walker time, $t$, reads (in $D=4$ space-time dimensions)~\cite{AEM1} (c.f. Appendix C for details):
\be
a(t)\propto t^{1+(\phi_0 + \tau_0/2)^{-1}}~, \qquad {\rm if}~~\phi_0 + \frac{\tau_0}{2} \ne 0~.
\label{scalerw}
\ee
The $\sigma$-model frame time, $x^0$, is related in this case to the cosmic time, $t$, by:
\be\label{x0}
x^0 = \sqrt{\ap}\left[\left(\phi_0+\frac{\tau_0}{2}\right)\frac{t-t_2}{\sqrt{A}}\right]^{-\left(\phi_0+\frac{\tau_0}{2}\right)^{-1}}~,
\ee
where $t_2$ represents and integration constant. Expanding Universes require $ 1+(\phi_0 + \tau_0/2)^{-1}> 0$, while absence of horizons, and hence a well-defined S-matrix,
occurs for
\be
\frac{\tau_0}{2} + \phi_0 \,>\, 0 ~.
\label{horizon}
\ee
The reader is reminded at this point that to ensure perturbative validity of our string tree-level considerations, the string coupling
\begin{equation}\label{string}
g_s = {\rm exp}(\phi)\propto \left(\frac{t}{\sqrt{A}}\right)^{-\frac{\phi_0}{\phi_0+\frac{\tau_0}{2}}}
\end{equation}
must be less than one asymptotically, for $t\rightarrow \infty$ pertaining to the exit phase. This implies $\phi_0 > 0$. In this way, we do not need to consider summation over
higher-order world-sheet genera, which is not known exactly in string $\sigma$-models (in fact, in Bosonic $\sigma$ models such a series is not Borel resummable~\cite{mende}).

Such Universes in general contain tachyons, and the important issue on the stability of the asymptotic
(after inflation) vacuum is in order. This issue can be examined in the redefined effective action
(\ref{einst}), in which the tachyon field ${\tilde T}$ has a diagonal kinetic term, and hence is considered as the ``physical'' mode. However, because this redefinition, as well as the action itself (\ref{einst}), involve the functions $f_i(T)$, the answer
depends on their precise form and hence the situation is inconclusive. Moreover, as we do not know the precise nature of the phase transition that leads to exit from the de Sitter phase in this case, there are additional important reasons for not being able to
say anything concrete about the asymptotic stability of the exit phase in the ``eternal inflation'' scenario, based on the condition (\ref{antialign}).

To overcome these uncertainties, in the next subsection we discuss a \emph{smooth interpolating} solution, which includes inflation at a certain stage and interpolates between flat Universes, corresponding to low-energy string effective field theories with well-defined S-matrix elements. Such an extension is useful in that it allows a discussion of string production, contributing to reheating at the end of the inflationary era.
Moreover, one can revisit in this context the issue of asymptotic stability of the ground state in the post inflationary phases in a more concrete way, as we now proceed to discuss.

\subsection{Closed-String Inflation as a ``short-time'' interpolating solution between power-law expanding Universes}

\begin{figure}[ht]
\centering
\includegraphics[width=12.6cm, height=8.4cm]{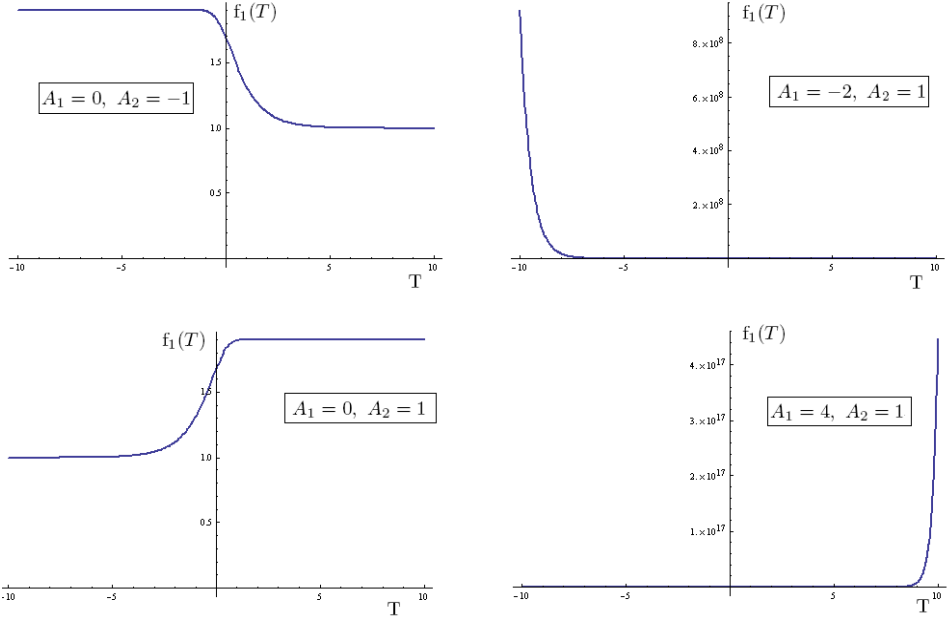}
\caption{{\footnotesize Plots of the function $f_1(T)=e^{A_1 T}\left[1+0.9\tanh(e^{A_2 T})\right]$, for various values of $A_1$ and $A_2$. In our work we are interested in solutions in which the function $f_1(T)$ asymptotes constant finite values, and $f_1'(T) \to 0$, as these correspond to asymptotic stable ground states after the exit from the inflationary phase. In  this sense, for our purposes here, we only keep the cases plotted on the left panel of the figure.}}
\label{new1}
\end{figure}

\begin{figure}[ht]
\centering
\includegraphics[width=12.6cm]{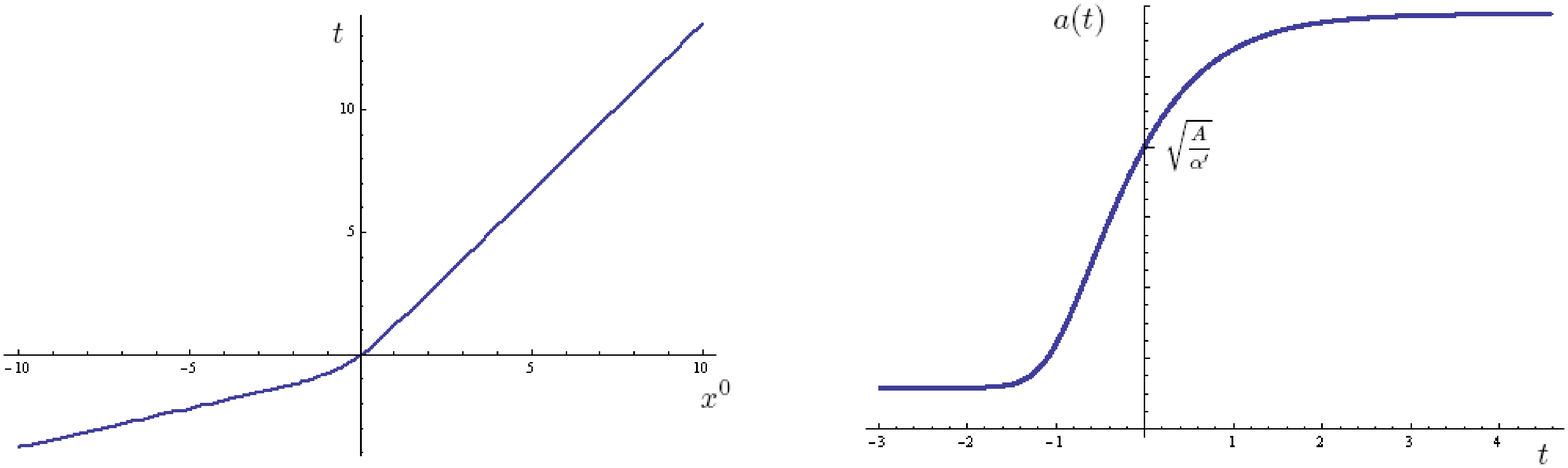} \vfill
\includegraphics[width=12.6cm]{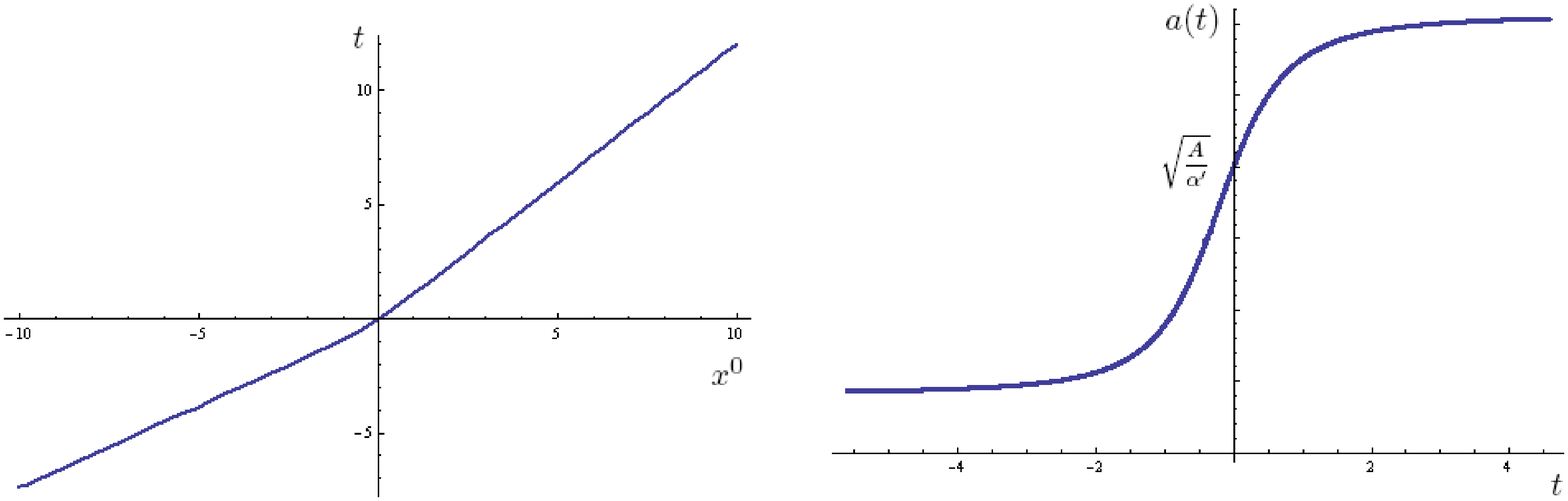}
\caption{{\footnotesize \emph{Upper}: Plot of the Einstein frame time, $t$, as a function of the sigma-model frame time, $x^0$, and of the scale factor, $a$, as a function of the cosmic time, $t$, for the choice for the function $f_1(T)=e^{A_1 T} \left[1+0.9\tanh(e^{A_2 T}) \right]$, and for field amplitudes satisfying $\tau_0=\frac{1}{A_2}$ and $2(1+\phi_0)=\frac{A_1}{A_2}$.
\emph{Lower:} Same plots of $t(x^0)$ and of $a(t)$, this time for the choice $f_1$: $e^{A_3 T} \left[1+0.5\frac{e^{A_4 T}}{\sqrt{e^{A_4 T}+1}} \right]$, and for field amplitudes satisfying $\tau_0=\frac{1}{A_4}$ and $2(1+\phi_0)=\frac{A_3}{A_4}$.
In all figures, $t$ is given in units of $\sqrt{A}$, while $\eta$ and $x^0$ are given in units of $\sqrt{\ap}$.}}
\label{new2}
\end{figure}

In the Einstein frame, and in terms of the cosmic Robertson-Walker time $t$,
defined through (\ref{ty}) of Appendix C,
our space-time metric configuration (\ref{configbis}) leads to a Friedman-Robertson-Walker (FRW) Universe and, as explained in Appendix C, we can determine the dependence of $t$ and the scale factor, $a$, on $\tilde x^0$. The string coordinate redefinition, $x^0\rightarrow  x^0\sim\exp(-\tilde x^0/\sqrt{A})$, makes it easier to see that under certain conditions (if
there is a cancellation between $\phi$ and $\ln f_1(T)$ terms in (\ref{wdetail})), the $\sigma$-model and the Einstein frame coincide, and one obtains the eternal inflation de Sitter solution of the previous subsection. If there is no such cancellation, different FRW universes are obtained, some of which might be characterized by well-defined asymptotic scattering matrix elements, and therefore be consistent solutions of perturbative string theory as well. It is this kind of solutions we are after in this section.

To study solutions for a general $f_1(T)$, we first write down the equation that relates the cosmic time, $t$, and the scale factor, $a$, in the Einstein frame, with the original $\sigma$-model frame time coordinate, $x^0$, in $D=4$:
\be
ds^2=dt^2-a^2(t)d{\bf r}^2 = e^\omega \left[\frac{A}{(x^0)^2}(dx^0)^2-\frac{A}{(x^0)^2}d{\bf x}^2 \right]
\ee
where $\omega$ is given by:
\be
\omega=-2\phi+\ln f_1(T)=-2\phi_0\ln \frac{x^0}{\sqrt{\ap}} +\ln f_1 \left(\tau_0 \ln \frac{x^0}{\sqrt{\ap}}\right)
\ee
This means that $t$ and $a$ are related  to $x^0$ by:
\bea \label{ta}
dt&=&\varepsilon(x^0) \frac{\sqrt{A}}{x^0} e^{\omega/2} dx^0 \\
a(t)&=&a_0 \frac{\sqrt{A}}{|x^0|} e^{\omega/2}
\eea
where $a_0$ is a dimensionless constant (that we will choose from now on to be equal to $1$) and $\varepsilon(x^0)$ can take the values $\pm1$ and it should take opposite values for negative and positive $x^0$, in order to ensure that $t(x^0)$ is a monotonic function. Let us assume that $\varepsilon(x^0)=+1$ for $x^0>0$ and $\varepsilon(x^0)=-1$ for $x^0<0$. A first remark from (\ref{ta}) is that the sigma-model time coordinate, $x^0$, plays the role of conformal time in the Einstein frame (independently of the exact form of $f_1(T)$):
\be
d\eta\equiv\frac{dt}{a(t)}=dx^0
\ee
Thus, although it is not possible to solve the above equations and obtain the scale factor $a$ as a function of $t$ analytically for any general function $f_1(T)$, it is straightforward to find the conformal scale factor, $C(\eta)$, defined as usually as:
\be
ds^2=dt^2-a^2(t) d{\bf r}^2\equiv C(\eta) \left(d\eta^2-d{\bf r}^2 \right)
\ee
Thus, substituting $\eta=x^0$ in (\ref{ta}), we get an analytic expression of $C(\eta)$, involving the function $f_1$ and the amplitudes $\phi_0$ and $\tau_0$.
\be \label{c}
C(\eta)~=~ \frac{A}{\ap}~ \left(\frac{\eta}{\sqrt{\ap}} \right)^{-2(1+\phi_0)} f_1\left(\tau_0\ln \frac{\eta}{\sqrt{\ap}}\right)
\ee
All the information about the geometry of the space-time in the Einstein frame is contained in this equation. Since there is a freedom in the choice of $f_1$, we may look at equation (\ref{c}) in the opposite way and try to find what form $f_1$ should take in order to give a specific scale factor, $C(\eta)$. From equation (\ref{c}) we find that the expression for $f_1(T)$ that will give the scale factor $C(\eta)$ is:
\be \label{f}
f_1(T)=\frac{\ap}{A}e^{\frac{2(1+\phi_0)}{\tau_0}T}~C \left(\sqrt{\ap} e^{T/\tau_0}\right)
\ee
We now show that we can obtain in this way the two forms of the conformal scale factor studied in \cite{gubser} as toy examples to discuss string production at the end of inflation.
These examples are physically interesting because they represent a universe with a limited time inflationary phase and flat space solutions in the infinite past and in the infinite future. Consider the following form for $f_1(T)$ (plotted in figure \ref{new1}) for various values of the constants $A_1$ and $A_2$ ($B$ and $\rho$ are positive, and $B \leq 1$) :
\be\label{f11}
f_1(T)=e^{A_1 T} \left[1+B\tanh(\rho e^{A_2 T}) \right]
\ee
One can numerically solve equations (\ref{ta}) to find that, if $\tau_0=\frac{1}{A_2}$ and $2(1+\phi_0)=\frac{A_1}{A_2}$, one obtains the dependence of the cosmic time $t$ on $x^0$ and the scale factor, $a(t)$ plotted in figure \ref{new2}. The conformal scale factor is of the form
\be\label{gubser1}
C(\eta)=\frac{A}{\ap}\left[1 + B\tanh(\rho \eta)\right]~,
\ee
which is plotted in figure \ref{new2}.

In the same way, the function:
\be\label{f12}
f_1(T)=e^{A_3 T} \left[1+B\frac{e^{A_4 T}}{\sqrt{e^{A_4 T}+\rho^2}} \right]~,
\ee
with $0<B<1$, gives us the conformal scale factor
\begin{equation}\label{gubser2}
C(\eta)=\frac{A}{\ap}\left[1+B\frac{\eta}{\sqrt{\eta^2+\rho^2}}\right]~,
\end{equation}
provided the amplitudes of the tachyon and dilaton satisfy: $\tau_0=\frac{1}{A_4}$ and $2(1+\phi_0)=\frac{A_3}{A_4}$, respectively. The corresponding dependence of the cosmic time on the sigma-model frame time, and the scale factor as a function of the cosmic time are plotted in figure \ref{new2}.

With these different choices for $f_1(T)$, we can now look again at the stability conditions for the model
in the post inflationary era, that is the behaviour of the tachyon terms in the effective target-space action (\ref{einstnd}). To this end, we first observe that the various choices (\ref{f11}), (\ref{f12}) of the function $f_1(T) > 0$, used above to derive physically interesting space times, correspond to appropriate field redefinitions of the tachyon field $T$, which do not depend on its space-time  derivatives and, hence, can be viewed as local field redefinitions that leave the scattering amplitudes invariant. Physically interesting models correspond to cases where $f_1 $ tends to a constant value asymptotically in cosmic time, which can be taken to be one. For and $f_1$ of the form (\ref{f11}), the appropriate choice of the constant $A_1$ in order for this to be true is $A_1=0$ (c.f. figure~\ref{new1}), which means that the dilaton amplitude has to be $\phi_0=-1$ in order to get the appropriate scale factor. As can be seen from figure~\ref{new1}, $f_1'(T) \to 0$ asymptotically in such a case as well. With these assumptions, target space-time physics in the Einstein frame at late times are described by the following effective action,
\bea\label{einstnd2}
S^{E}_{\mbox{\footnotesize late times}}&\sim&\int d^D x\sqrt{-g} ~\Bigg\{~ R  + e^{\frac{4\phi}{D-2}} \left[\frac{D-26}{6\ap} + f_0(T)\right] -\left[\frac{4(D-1)}{D-2} - 4f_2(T) \right]\partial \phi \cdot \partial \phi\nn
&&~~~~~~~~~~~~~~~~~ - f_3(T) \partial T \cdot \partial T  -f_4(T) \partial\phi\cdot \partial T ~\Bigg\} ~,
\eea
The time coordinate is proportional to the $\sigma$-model frame time: $t=\sqrt{\frac{A}{\ap}}x^0$, and the dilaton and tachyon configurations are $\phi=-\ln \left(t/\sqrt{A}\right)$ and $T=\tau_0 \ln \left(t/\sqrt{A}\right)$. This means that the prefactor of the potential terms in the above action falls to zero like $t^{-2}$ and all the kinetic terms, $\partial\phi\cdot\partial\phi$, $\partial\phi\cdot\partial T$ and $\partial T\cdot\partial T$ also fall like $t^{-2}$. We can therefore place a mild requirement on the polynomial functions $f_2(T)$, $f_3(T)$ and $f_4(T)$ (possibly milder than (\ref{restrictions})): for large $t$ they are such that the tachyon and dilaton derivative terms do not diverge, but tend to constant values. For example, for a tachyon field with amplitude $\tau_0=1$, the requirement $f_i(T)<Ae^{2T}$ ($i=2,3,4$) for $T\rightarrow\infty$, suffices. In this way, the cosmological instabilities disappear and one is left with a Minkowski Universe in the physically interesting cases. Note also that in this case when looking at the string coupling $g_s$ (c.f. \ref{string}) falls asymptotically to zero for large positive $x^0$, which now corresponds to large positive cosmic time $t$. Thus, we do not need to sum over higher-order world-sheet genera in this case either. Similar conditions hold for the choice (\ref{f12}) for $f_1$.

As a final remark, we note that the analysis in this subsection can
be generalized to higher than four target-space dimensions. Indeed, in such a case
the conformal scale factor reads: $C(\eta)\propto\left(\eta/\sqrt{\ap}\right)^{-2-\frac{4\phi_0}{D-2}}\left[f_1\left(\tau_0\ln (\eta/\sqrt{\ap})\right)\right]^{\frac{2}{D-2}}$ and the class of functions $f_1$ that will lead to space-time cosmologies
with intermediate inflationary periods will again be combinations of exponentials $e^T$, but of different form than (\ref{f11}) and (\ref{f12}). This is due to the different conditions that the tachyon and dilaton amplitudes satisfy. Given that
$\phi_0=-\frac{D-2}{2}$, a plausible form for $f_1(T)$ is: $f_1(T)\propto \left[C(e^{bT})\right]^{\frac{D-2}{2}}$, with $b$ a constant. This leads to a conformal scale factor $C(\eta)$, provided the tachyon amplitude is $\tau_0=b^{-1}$. For the conformal factors discussed in this subsection, all these functions $f_1$ will asymptotically fall to a constant value, for large cosmic times. Thus, the above-mentioned asymptotic stability conditions, in the sense of decaying tachyon effects in
the effective action, are still valid for the higher-dimensional case.

\subsection{String production at the end of inflationary era in the interpolating Solution}

The scaling solutions (\ref{gubser1}) and (\ref{gubser2}) have been used as toy examples in the analysis of \cite{gubser} for a discussion of string production and (p)reheating in inflationary scenarios within string theory. However, no specific attempt had been made there to link such toy examples to realistic string cosmologies. As we have seen above, both types of solutions can indeed arise in the context of string cosmologies, provided both tachyon and dilaton fields are taken into account.

In fact, it is important to realize that the form of the interpolating solution, which includes inflation for a relatively short period of time in the Universe's evolution, depends crucially on the form of the tachyon form-factor function $f_1(T)$. The form of the latter does not affect the perturbative string scattering amplitudes, in the sense that its form can be changed by local field redefinitions.
The only restriction is the absence of cosmic horizons, so that scattering amplitudes are well defined.

In this sense, one may discuss string production and (p)reheating in our scenarios following  the analysis of \cite{gubser}, where we refer the interested reader for details.
The issues that are of interest to us here refer to
the conditions for preheating, namely the possibility that coherent oscillations of the inflaton field
towards the end of inflation are responsible for oscillating mass terms of other bosonic fields appearing in general in string theory.
In higher-dimensional strings, there are the moduli fields which enter the game. This is not true, however, in our four-dimensional ($D=4$) scenarios, where we only have two kinds of fields and no extra dimensions; hence the conditions for preheating have to be rethought, as compared to the higher-dimensional analysis of \cite{gubser}. We shall come back to these important physical applications of our scenario
in a forthcoming publication.

\section{Conclusions and Outlook \label{sec:4}}

In this paper we have discussed the emergence of inflation in a simple bosonic stringy sigma model, living in a four-dimensional target space, which as we have seen can be obtained as a consistent Weyl invariant solution to all orders in the Regge slope, $\alpha '$, provided non-trivial dilaton, tachyon and metric backgrounds are present. In fact, the presence of both tachyon and dilaton backgrounds is essential for obtaining inflation. Specifically, the tachyonic background, which describes an initial cosmological instability, during the inflationary epoch of the string Universe cancels the effects of the dilaton in a way that we have explained precisely above. We have discussed two classes of inflationary solutions: (i) a de Sitter Universe with ``eternal'' acceleration, which cannot characterize perturbative strings due to cosmic horizons, and in which the exit from the inflationary phase may be provided by a (yet unknown) phase transition mechanism, external to our mathematical solution, and (ii) a smooth solution,
 whereby inflation spans a short period in the history of the string Universe, interpolating between asymptotically flat Minkowski Universes.
There are cases in this latter class of solutions in which the tachyonic instabilities disappear asymptotically in cosmic times, and hence such cases are of physical interest, as they can serve as self consistent mathematical models for smooth exit from inflation and prototypes for discussions on (p)reheating. Such interpolating scenarios can characterize perturbative string theory in the sense of corresponding to well-defined scattering amplitudes.

The main difference between our approach and others in the literature (see eg. \cite{swanson}), apart from the $\alpha '$-resummation nature of the solution, and its four-space-time dimensional character, is that we do not require consistence of the generic effective spacetime action, (\ref{Sigma}) with the standard perturbative vacuum (flat target spacetime and linear dilaton). In fact, we have found a configuration for the background fields that is exact in $\ap$ and is not smoothly connected with the perturbative vacuum.

This is why we do not have to impose any further restrictions to the functions $f_i(T)$ that appear in (\ref{Sigma}), other than conditions like (\ref{restrictions}) that ensure that there are no ghost fields and that the model is stable. The freedom in  the choice of $f_1(T)$ that we are left with can lead, as we saw, to different target spacetime cosmologies, including universes which expand exponentially forever, or for a limited time. This freedom is missing if one demands connection with the perturbative vacuum and the only exact solution for the tachyon field found so far for this case is the null tachyon, $e^{\beta\cdot X}$, with $\beta$ a null vector.

This solution is not of interest to our approach, since we demand that our fields depend only on the time-coordinate of the sigma-model frame, $X^0$. In this approach, $f_1(T)$ is taken to be $1-T^2+{\cal O}(T^3)$, which does not fit with our approach as we don't demand connection with the perturbative vacuum, and indeed our solution for $T$ can take values much greater than $T=0$. In reference \cite{yang_zwiebach}, a logarithmic (in $\sigma$-model-frame time) solution for the dilaton and tachyon fields is also found, but to lowest order in $\ap$. However, there the function $f_1(T)$ is taken to be equal to $1$, thus a different FRW universe is found in the Einstein frame, namely one which is lead into a ``big crunch''.

There are many important physical properties of our inflationary solutions that we have not examined here, but we intend to come back to them in a forthcoming work. First of all, the detailed mechanism for string production at the end of inflation and its effects on the (p)reheating of the string Universe. Second, the possible appearance of non Gaussianities as a result of the existence of two scalar modes (dilaton, tachyon) in the spectrum. Last but not least, issues associated with the r\^ole of the asymptotically non constant dilatons and their effects on the running with cosmic time of the various fundamental ``constants'' of nature, such as gauge couplings and gravitational constants, which may kill the model altogether if the associated variations at late eras of the Universe are not within the currently accepted experimental bounds.
\emph{Affaire \`a suivre}...

\section*{Acknowledgements}

The work of A.K. and N.E.M. is supported by the European Union through the FP6 Marie Curie Research and Training Network \emph{UniverseNet} (MRTN-CT-2006-035863). A.K. also wishes to thank the
 Alevizatos-Kontos Foundation of the University of Athens (Greece) for additional partial support through a study-abroad scholarship.

\section*{Appendix A: $\alpha'$-resummed evolution equation}

We start with the bare Euclidean action defined on a flat world sheet:
\be
S_\sigma=\frac{1}{4\pi\ap}\olo \xi \left\{\delta^{ab} g_{\mu\nu}(\tx^0) \partial_a \tx^{\mu}\partial_b\tx^{\nu}
+4\pi\alpha^{'}\lambda\Lambda^2 \left(\tx^0\right)^2\right\},
\ee
where $\lambda$ is our running parameter.
The bare fields are denoted by $\tx^\mu$ to distinguish them from the classical fields that will be defined shortly.
If $V^\mu$ are source for the fields $\tx^\mu$, the partition function is (for a fixed world sheet metric):
\be
Z=\int{\cal D}[\tx] e^{-S - \int V_\nu \tx^\nu} \equiv e^{-W}
\ee
where $W$ is the connected graph generator functional, and the classical fields are
\be
X^\mu = <\tx^\mu> = \frac{1}{Z} \int{\cal D}[\tx]\tx^\mu e^{-S-\int V_\nu \tx^\nu}=\frac{\delta W}{\delta V_\mu}.
\ee
Note that the second functional derivatives of $W$ are then
\be
\frac{\delta^2 W}{\delta V_\mu (\xi)\delta V_\nu (\zeta)} = X^\mu (\xi) X^\nu (\zeta) - <\tx^\mu(\xi) \tx^\nu(\zeta)>.
\ee
The proper graphs generator functional $\Gamma$ is introduced as the Legendre transform of $W$:
\be
\Gamma = W - \olo \xi V^\mu X_\mu,
\ee
and satisfies
\bea
\frac{\delta \Gamma}{\delta X^\mu}&=&-V_\mu\nn
\partial_\lambda\Gamma&=&\partial_\lambda W.
\eea
The second derivatives of $\Gamma$ and $W$ are related by
\be\label{gammaW}
\frac{\delta^2 \Gamma}{\delta X^\mu (\xi) \delta X^\nu (\zeta)}
=-\frac{\delta V_\nu(\zeta)}{\delta X^\mu (\xi)}
=-\left(\frac{\delta^2 W}{\delta V_\mu (\xi) V_\nu (\zeta)}\right) ^{-1}.
\ee
We are now interested in finding the evolution of $\Gamma$ with respect to $\lambda$.
We have (a dot over a letter denotes a derivative with respect to $\lambda$)
\bea
\dot W&=&\Lambda^2 \int d^2\xi\left<(\tx^0)^2\right>\nn
&=&\Lambda^2\int d^2\xi(X^0)^2-\Lambda^2\mbox{ Tr} \left\{\frac{\delta^2 W}{\delta V_0 \delta V_0} \right\},
\eea
such that, taking into account eq.(\ref{gammaW}), we find the following self consistent evolution equation for $\Gamma$:
\be \label{yioupi}
\dot{\Gamma} = \Lambda^2\olo \xi \left(X^0\right)^2 +
\la \mbox{ Tr} \left\{ \left(\frac{\delta^2 \Gamma}{\delta X^0\delta X^0}\right)^{-1} \right\}.
\ee
Note the important difference with Wilsonian exact renormalization equations: although the trace appearing in
eq.(\ref{yioupi}) needs to be regularized, our approach makes use of the
{\it fixed} world sheet cut off $\Lambda$, and the running parameter is $\lambda$.\\
In order to solve this evolution equation,
we will assume, in the framework of the gradient expansion, the following functional dependence of the effective action on $X$:
\be
\Gamma[X]=\frac{1}{4\pi\alpha'} \olo \xi \left\{\delta^{ab} \frac{A}{(X^0)^2}\eta_{\mu\nu}\partial_a X^\mu \partial_b X^\nu
+4\pi\alpha'\la T(X^0)\right\},
\ee
where $\la$ is the world sheet UV cut-off.  The corresponding evolution equation for the tachyon is then
\be \label{final}
\dot T(X^0) = (X^0)^2 + \alpha'\frac{(X^0)^2}{2A} \ln \left[1+\frac{A}{2\pi\alpha'(X^0)^2 T''(X^0) }\right]
\ee

\section*{Appendix B: Field redefinitions}

If we start from a theory with $\beta$-functions $\beta^i$ corresponding to the various background fields, $g^i$, and we make a general field redefinition $g^i \rightarrow \tilde{g}^i \equiv g^i + \delta g^i$, the  $\beta^i$'s transform in the following way \cite{metsaev}:
\be
\beta^i \rightarrow \tilde{\beta}^i = \beta^i + \delta g^j \frac{\delta \beta^i}{\delta g^j} -  \beta^j \frac{\delta(\delta g^i)}{\delta g^j}
\ee
\nin For our background fields, $g_{\mu\nu}$, $\phi$ and $T$, and for a time-dependent configuration, this means that the general field redefinition:
\bea
g^{\mu\nu} & \rightarrow & \tilde{g}^{\mu\nu} \left(g^{\mu\nu}, \phi, T \right) \nn
\phi & \rightarrow & \tilde{\phi}\left(g^{\mu\nu}, \phi, T \right) \nn
T & \rightarrow & \tilde{T}\left(g^{\mu\nu}, \phi, T \right)
\eea
\nin leads to the following change in the $\beta$-functions:
\bea\label{transform}
\tilde{\beta}^i (X^0)- \beta^i(X^0) &=& \int dY^0 \frac{\delta \beta^i(X^0)}{\delta g^j(Y^0)} \left( \tilde{g}^j (Y^0)- g^j(Y^0) \right) \nn
&&- \int dY^0  \beta^j (Y^0) \frac{\delta \left( \tilde{g}^i (X^0)- g^i(X^0) \right)}{\delta g^j(Y^0)}
\eea

Using this property of the $\beta$-functions, and the fact that our theory is invariant under general field redefinitions, we can make the one-loop $\beta$-functions (\ref{1loop}) vanish, by applying the field redefinition (\ref{redef}):
\bea \label{redef2}
g_{\mu\nu} &\rightarrow& \tilde{g}_{\mu\nu} = g_{\mu\nu} + \ap g_{\mu\nu}
\Big(a_1 R + a_2 \partial \phi \cdot \partial \phi + a_3 \partial \phi \cdot \partial T\nn
&&~~~~~~~~~~~~~~~~~~~~~~~~+a_4 \partial T\cdot \partial T + a_5 \nabla^2\phi +a_6\nabla^2 T\Big)  \nn
\phi &\rightarrow& \tilde{\phi} = \phi + \ap
\Big(b_1 R + b_2 \partial \phi \cdot \partial \phi + b_3 \partial \phi \cdot \partial T\nn
&&~~~~~~~~~~~~~~~~~~~~+b_4 \partial T\cdot \partial T + b_5 \nabla^2\phi +b_6\nabla^2 T\Big)  \nn
T &\rightarrow& \tilde{T} = T+\ap a R T+\ap
\Big(c_1 R + c_2 \partial \phi \cdot \partial \phi + c_3 \partial \phi \cdot \partial T\nn
&&~~~~~~~~~~~~~~~~~~~~~~~~~~+c_4 \partial T\cdot \partial T + c_5 \nabla^2\phi +c_6\nabla^2 T\Big),
\eea
\nin where $a$, $a_1$ ... $a_6$, $b_1$ ... $b_6$ and $c_1$ ... $c_6$ are all dimensionless constants that we are free to choose. This redefinition doesn't change the $X^0$-dependence of the fields, as $\delta g_{\mu\nu}$ is proportional to $g_{\mu\nu}$, $\phi$ is just shifted by a constant and $T$ is shifted by a constant and a term proportional to itself. This is very important, as in the main text we showed that for our configuration (\ref{config}) all the $\beta$-functions have homogeneous dependence on $X^0$, to all orders in $\ap$, besides $\beta^T$ which contains one inhomogeneous term: $-2T$. Now we will show that the homogeneity of $\beta^{g}_{\mu\nu}$ and $\beta^\phi$ is not affected by the above redefinition. Thus, as long as $g_{\mu\nu}$ remains proportional to $(X^0)^{-2}\eta_{\mu\nu}$ and $\phi$ and $T$ remain proportional to $\ln\left(\frac{X^0}{\sqrt{\alpha'}}\right)$ (plus constant shifts), $\beta^{g}_{\mu\nu}$ and $\beta^\phi$ remain homogeneous. As for $\beta^T$, the only term that affects its homogeneity in the above redefinition is the linear term $\ap a RT$, which as we will show adds a new linear term to the $\beta$-function; this new term, upon appropriate choice of the constant $a$, can lead to the cancellation of the original linear term. Every other new or old contribution to the $\beta^T$ is homogeneous (i.e. for our field configuration it is a constant). Thus, after performing this field redefinition, all three $\beta$-functions are homogeneous and contain in them the 18 new constants, $a_i$, $b_i$, $c_i$, that we are free to choose in a way to make all $\beta$-functions equal to zero.

We will study in more detail the effect of the above redefinition in two steps. First we will present the change that the term $\ap aRT$ in $\delta T$ causes to the three $\beta$-functions: it makes them all homogeneous; and then we will shortly present the effect of all the other terms and show that we have enough free parameters to make all $\beta$-functions vanish (to second order in $\ap$).

Let's start with $\ap aRT \equiv (\delta T)_1$: First, note that $\beta^\phi$ has no dependence on $T$ and thus is not affected by this term (or any other term in $\delta T$). The graviton and the tachyon $\beta$-functions get contributions by the following new terms:
\be
\left(\delta \beta^{g}_{\mu\nu}\right)_1 (X^0) = \int dY^0 (\delta T)_1(Y^0) \frac{\delta \beta^{g}_{\mu\nu}(X^0)}{\delta T(Y^0)},
\ee
\bea\label{deltabetat}
\left(\delta \beta^T\right)_1 (X^0) &=& \int dY^0 (\delta T)_1(Y^0) \frac{\delta \beta^T(X^0)}{\delta T(Y^0)} - \int dY^0 \beta^{g}_{\mu\nu}(Y^0) \frac{\delta( (\delta T)_1(X^0))}{\delta g_{\mu\nu} (Y^0)}\nn
&&- \int dY^0 \beta^T(Y^0) \frac{\delta( (\delta T)_1(X^0))}{\delta T(Y^0)}
\eea
\nin The change in the graviton $\beta$-function does not affect its homogeneity, as it only yields the terms:
\be
\left(\delta \beta^{g}_{\mu\nu}\right)_1 (X^0) = - 2(\ap)^2 a R \partial_\mu T \partial_\nu T -(\ap)^2 aT \left(\partial_\mu R \partial_\nu T + \partial_\mu T \partial_\nu R \right)
\ee
\nin For our configuration, $R$ is given by a constant, so only the first of the above terms survive, and this is homogeneous to the rest of the terms in $\beta^{g}_{\mu\nu}$:
\bea \label{deltabetag1}
\left(\delta \beta^{g}_{00}\right)_1 (X^0) &=& -2 a \frac{(\ap)^2 (3D-D^2-2)}{A}\frac{\tau_{0}^{2}}{(X^0)^2}\nn
\left(\delta \beta^{g}_{ij}\right)_1 (X^0) &=& 0
\eea

The change in the tachyon $\beta$-function yields terms that are linear in the tachyon field. By substituting the following expressions in (\ref{deltabetat}),
\bea
\frac{\delta \beta^T(X^0)}{\delta T(Y^0)} &=& -2\delta(X^0-Y^0)-\frac{\ap}{2}\nabla^2 \delta(X^0-Y^0)+\ap \partial^\mu\phi\partial_\mu \delta(X^0-Y^0) \nn
\frac{\delta( (\delta T)_1(X^0))}{\delta g_{\mu\nu} (Y^0)} &=&  \ap aT \left[R^{\mu\nu}\delta(X^0-Y^0)+\nabla^\mu\nabla^\nu \delta(X^0-Y^0) - g^{\mu\nu} \nabla^2 \delta(X^0-Y^0) \right]\nn
\frac{\delta( (\delta T)_1(X^0))}{\delta T(Y^0)} &=& \ap a R \delta(X^0-Y^0)~,
\eea
\nin we are lead to:
\be
\left(\delta \beta^T\right)_1 = \ap a T \left[g^{\mu\nu}\nabla^2\beta^{g}_{\mu\nu} - R^{\mu\nu}\beta^{g}_{\mu\nu} - \nabla^\mu\nabla^\nu\beta^{g}_{\mu\nu} \right]+ (\ap)^2 a T \partial^\mu\phi\partial_\mu R
\ee
\nin For our configuration, the second term is equal to zero, whereas the first term is just the tachyon field, $T$, multiplied by a constant:
\be
\left(\delta \beta^T\right)_1 = a \frac{(\ap)^2}{A^2} (D-1) \left[D-D^2 -4(D-1)\phi_0 +(D-2)\tau_{0}^{2}\right] T
\ee

\nin Thus, by making the choice
\be
a=\frac{2A^2}{(\ap)^2} \frac{1}{ (D-1) \left[D-D^2 -4(D-1)\phi_0 +(D-2)\tau_{0}^{2}\right]},
\ee
\nin this new term in $\beta^T$ cancels the original linear term, $-2T$, and thus leaves us with all $\beta$-functions homogeneous. Note also that after fixing $a$ the contribution of the term $\ap aRT$ to $\beta^{g}_{\mu\nu}$ is (\ref{deltabetag1}):
\bea
\left(\delta \beta^{g}_{00}\right)_1 (X^0) &=& \frac{A}{(X^0)^2} \frac{(D-2) \tau_{0}^{2}}{D-D^2 -4(D-1)\phi_0 +(D-2)\tau_{0}^{2}} \nn
\left(\delta \beta^{g}_{ij}\right)_1 (X^0) &=& 0
\eea

One can check that all the other terms that appear in the redefinition (\ref{redef2}) will be homogeneous to the already existing terms in the $\beta$-functions: the new terms in $\beta^{g}_{\mu\nu}$ will be proportional to $(X^0)^{-2}$ and the new terms in $\beta^\phi$ and $\beta^T$ will be constants. Take for example the term $(\delta \phi)_3 \equiv \ap b_3 \partial \phi \cdot \partial T$ which affects all three $\beta$-functions. Its contribution to each $\beta$-function is given by:
\bea
\left(\delta\beta^{g}_{\mu\nu}\right)_3&=& \int(\delta\phi)_3\frac{\delta\beta^{g}_{\mu\nu}}{\delta \phi}=2\ap \nabla_\mu\nabla_\nu (\delta\phi)_3 \\
\left(\delta\beta^\phi \right)_3&=&  \int(\delta\phi)_3\frac{\delta\beta^\phi}{\delta \phi} - \int \beta^{g}_{\mu\nu}\frac{\delta(\delta\phi)_3}{\delta g_{\mu\nu}}- \int \beta^\phi \frac{\delta(\delta\phi)_3}{\delta\phi} - \int \beta^T \frac{\delta(\delta\phi)_3}{\delta T}\nn
&=& -\frac{\ap}{2}\nabla^2 (\delta\phi)_3 +2\ap\partial\phi\cdot\partial(\delta\phi)_3 + b_3 g^{\kappa\mu}g^{\lambda\nu}\beta^{g}_{\kappa\lambda} \partial_\mu\phi\partial_\nu T \nn
&& -b_3\partial T \cdot \partial \beta^\phi - b_3\partial \phi \cdot \partial \beta^T \\
\left(\delta\beta^T \right)_3&=& \int (\delta\phi)_3 \frac{\delta\beta^T}{\delta\phi} = \ap \partial T \cdot \partial (\delta\phi)_3
\eea
\nin For our field configuration, this gives $\left(\delta\beta^{g}_{\mu\nu}\right)_3=\left(\delta\beta^T\right)_3=0$ and $\left(\delta\beta^\phi \right)_3=-b_3 \tau_{0}^{2} \frac{(\ap)^2}{A^2} (D-1+\tau_0^2) $. Similarly, all other terms in (\ref{redef2}) give contributions to one or more of the $\beta$-functions; each contribution is proportional to one of the 18 free parameters $(a_i,b_i,c_i)$ and is homogeneous in $X^0$ to the rest of the $\beta$-function. Taking into account the original two-loop $\beta$-functions, what we get in the end is:
\bea
\tilde\beta^{g}_{00} &=& \frac{\tilde{E}_1}{(X^0)^2} +{\cal O} \left({\ap}^3\right) \nn
\tilde\beta^{g}_{ij} &=& \frac{\tilde{E}_2}{\delta_{ij}}{(X^0)^2} +{\cal O} \left({\ap}^3\right) \nn
\tilde\beta^\phi &=& \tilde{E}_3 +{\cal O} \left({\ap}^3\right) \nn
\tilde\beta^T &=& \tilde{E}_4 +{\cal O} \left({\ap}^3\right)
\eea
\nin with $\tilde{E}_1$, $\tilde{E}_2$, $\tilde{E}_3$ and $\tilde{E}_4$ being constants which are formed by linear combinations of the 18 parameters $(a_i,b_i,c_i)$; the parameters $a$, $A$, $\phi_0$, $\tau_0$ and $D$ will appear in these expressions, and everything will be of order $(\ap)^2$. In order to make the $\beta$-functions vanish, we just have to choose $(a_i,b_i,c_i)$ such that $\tilde{E}_1=\tilde{E}_2=\tilde{E}_3=\tilde{E}_4=0$ and it is evident that we have enough freedom to do this.

\section*{Appendix C: Einstein frame in the presence of closed-string tachyon backgrounds}

We commence our discussion from the scalar-curvature part of the effective action for the graviton
, $g$, dilaton, $\phi$ and tachyon $T$ backgrounds in the $\sigma$-model frame~\cite{aben}:
\be \label{act}
S_\sigma\sim \int d^Dx \sqrt{-g}e^{-2\phi}\left[f_1(T) R^\sigma + \dots\right],
\ee
where $f_1(T)$ is a function of $T$.\\
In order to pass to the Einstein frame, where the pre-factor of $R$ in (\ref{act}) is unity,
we have to make a redefinition of the metric of the following form~\cite{aben}:
\be \label{red}
g_{\mu\nu} \rightarrow g^E_{\mu\nu}=e^{\omega(\phi(x),T(x))}g_{\mu\nu}
\ee

\nin The action in the Einstein frame should be of the form (quantities in this frame are indicated
here with the subscript $E$ for concreteness):
\be \label{ein}
S \sim \int d^Dx \sqrt{-g^E}\left[ R^E + \dots\right]
\ee
Now, with the redefinition (\ref{red}) we have:
\bea
\sqrt{-g}&=&\sqrt{-g^E}\exp\left(-\frac{Dw}{2}\right) \\
R^\sigma &=& e^\omega \left[ R^E + (D-1)\nabla_{E}^{2}\omega-\frac{(D-1)(D-2)}{4}(\partial \omega \cdot \partial \omega)^E \right]
\label{ricci}
\eea
where by $\nabla_E$ we mean here the covariant derivative defined in the metric $g^{E}_{\mu\nu}$, and similarly $(\partial \omega \cdot \partial \omega)^E \equiv g^{(E)\mu\nu} \partial_\mu \omega \partial_\nu\omega$. So, to obtain the required form, (\ref{ein}), we must have:
\bea\label{wdetail}
1&=&\exp \left(-\frac{D\omega}{2} +\omega-2\phi \right) f_1(T)\nn
\mbox{or}~~~~\omega(\phi, T) &=& \frac{-4\phi+2\ln f_1(T)}{D-2}.
\eea
Thus, the metric redefinition that allows the passage to the Einstein frame is:
\be \label{re}
g_{\mu\nu} \rightarrow g^E_{\mu\nu}=e^{-\frac{4\phi}{D-2}} \left[f_1(T)\right]^{\frac{2}{D-2}}g_{\mu\nu}~.
\ee
Our metric configuration (\ref{config}) in the $\sigma$-model frame is:
\be
g_{\mu\nu}=\frac{A}{(x^0)^2} \eta_{\mu\nu}=\frac{A}{(x^0)^2}\mbox{diag} (1,-1,-1,-1)
\ee
This is an inflationary Robertson-Walker universe, with $x^0$ playing the role of conformal time:
\be
ds^2=\frac{A}{(x^0)^2}\left[(dx^0)^2-(d {\bf x})^2\right]
\ee
We pass to the Einstein frame, by making the metric redefinition (\ref{re}):
\be
ds^2 \equiv dt^2 - a^2(t) (d {\bf r})^2 =  \frac{A e^\omega}{(x^0)^2}\left[(dx^0)^2-(d {\bf x})^2\right]
\ee
In this frame, $x^0$ still has the role of conformal time, but now with a different conformal scale factor. The cosmic time $t$ is defined through~\cite{aben}:
\bea\label{ty}
dt &=& \varepsilon~e^{\omega/2}\frac{\sqrt{A}}{x^0}dx^0 \nn
\varepsilon &=& \pm 1
\eea
and
\be\label{at}
a(t) \sim \frac{\sqrt{A}}{|x^0|}e^{\omega/2} =\frac{\sqrt{A}}{|x^0(t)|}\exp\left(\frac{- 2\phi (t) + \ln f_1(T(t))}{D-2} \right)
\ee
For our configuration for the dilaton and tachyon fields (\ref{configbis}):
\bea\label{conf2}
\phi(x^0)&=&\phi_0 \ln(x^0/\sqrt{\alpha'}) \nonumber \\
T(x^0)&=&\tau_0 \ln(x^0/\sqrt{\alpha'})~,
\eea
and the choice $f_1(T) = e^{-T}$ used in \cite{tseytlin} and here, we thus have the following possibilities:

\begin{itemize}

\item  $2\phi_0 + \tau_0 = 0$, in which case $e^{\omega/2}=1$ and the Einstein frame coincides with the $\sigma$-model frame. From (\ref{ty}) we obtain:
$$
t = t_1 + \varepsilon \sqrt{A}\ln\frac{x^0}{\sqrt{\ap}},
$$
with $t_1$ an integration constant, and thus from (\ref{at}) we obtain inflation (as in the $\sigma$-model frame):
\begin{equation}
a (t) = a_0 \exp\left(-\varepsilon~ \frac{t-t_1}{\sqrt{A}}\right)
\label{inflsol2}
\end{equation}
provided $\varepsilon = - 1$~. This solution is characterized by a Hubble parameter:
$$ H_I = \frac{1}{\sqrt{A}}~.$$

\item $2\phi_0 + \tau_0 \ne 0$, and from (\ref{ty}) we obtain:
\bea \label{ta2}
t - t_2 = -\varepsilon \frac{\sqrt{A}(D-2)}{2\phi_0 + \tau_0}\left(\frac{x^0}{\sqrt{\ap}}\right)^{-\frac{2\phi_0+\tau_0}{D-2}}
\eea
where $t_2$ is an integration constant. In this case, for asymptotically long times
$t$ both the dilaton and tachyon configurations (\ref{conf2}) scale with $t$ as
\begin{eqnarray}
\phi(t) &\to & -(D-2)\frac{\phi_0}{2\phi_0 + \tau_0} \ln\frac{t}{\sqrt{\ap}}~, \nonumber \\
T(t) & \to & -(D-2)\frac{\tau_1}{2\phi_1 + \tau_1} \ln\frac{t}{\sqrt{\ap}}~,
\label{asympt2}
\end{eqnarray}
for large cosmic Robertson-Walker times $t \gg \sqrt{\alpha '}$.

Upon choosing
\be
\label{signe}
2\phi_0 + \tau_0 \, > \, 0~~~~~~~~~~~~~~~\mbox{and}~~~~~~~~\varepsilon \, = \, -1~,
\ee
we obtain a power-law \emph{expanding} universe:
\bea\label{expand}
a(t)  =a_0 \left(\frac{t - t_2}{\sqrt{A}}\right)^{1 + \frac{D-2}{2\phi_0 + \tau_0}}~.
\eea

This universe is characterized by the \emph{absence} of \emph{cosmic horizons}, and hence has well-defined S-matrix elements.

\end{itemize}

\end{document}